\documentclass[12pt, draftclsnofoot, onecolumn]{IEEEtran}
\ifCLASSINFOpdf
   \usepackage[pdftex]{graphicx}
  \DeclareGraphicsExtensions{.pdf,.jpeg,.png}
\else
\fi

\usepackage{amssymb,amsmath,amsthm}
\usepackage{epstopdf}
\usepackage{color}
\usepackage{cite}
\usepackage{caption}
\usepackage{subcaption}
\usepackage{empheq}

\usepackage{float}

\usepackage{tikz}
\usetikzlibrary{automata,arrows,positioning,calc,decorations.pathreplacing,decorations.markings,shapes.geometric}

\newtheorem{remark}{Remark}

\newcommand{\GD}{\Gamma_{\text{D2D}}}
\newcommand{\GB}{\Gamma_{\text{BS}}}

\tikzset{
    >=stealth',
    naming/.style={align=center,font=\small},
    pil/.style={->,shorten <=2pt,shorten >=2pt},
    pil_rev/.style={<-,shorten <=2pt,shorten >=2pt},
    pil_dash/.style={->, dashed,shorten <=2pt,shorten >=2pt},
    station/.style={	naming,draw,shape=dart,shape border rotate=90, minimum width=10mm, minimum height=10mm,outer sep=0pt,inner sep=3pt},
    antenna/.style={insert path={-- coordinate (ant#1) ++(0,0.25) -- +(135:0.25) + (0,0) -- +(45:0.25)}}
}

\newcommand{\BS}[1]{%
\begin{tikzpicture}
\node[station] (base) {#1};
\draw[line join=bevel] (base.100) -- (base.80) -- (base.110) -- (base.70) -- (base.north west) -- (base.north east);
\draw[line join=bevel] (base.100) -- (base.70) (base.110) -- (base.north east);
\draw[line cap=rect] ([yshift=0pt]base.north) [antenna=1];
\end{tikzpicture}
}

\begin{document}
\title{Coded Caching Clusters with Device-to-Device Communications}
\author{Joonas~P\"a\"akk\"onen, Amaro~Barreal, Camilla~Hollanti,~\IEEEmembership{Member,~IEEE}, and~Olav~Tirkkonen,~\IEEEmembership{Member,~IEEE}
\thanks{J. P\"a\"akk\"onen and O. Tirkkonen are with the Department of Communications and Networking, Aalto University, Finland. A. Barreal and C. Hollanti are with the Department of Mathematics and Systems Analysis, Aalto University, Finland. (Emails: firstname.lastname@aalto.fi)}
\thanks{The authors are financially supported by the Academy of Finland under Grants \#276031, \#282938, \#283262 and \#284725, the Finnish Funding Agency for Technology and Innovations under grant 2383/31/2014, as well as a grant from the Finnish Foundation for Technology Promotion. The support from the ESF COST Action IC1104 is gratefully acknowledged.}
\thanks{Parts of this work were presented at IEEE GLOBECOM 2013 \cite{globecom} and MACOM 2015 \cite{macom}.}}

\maketitle

\begin{abstract}
We consider a geographically constrained caching community where popular data files are cached on mobile terminals and distributed through Device-to-Device (D2D) communications. Further, to ensure availability, data files are protected against user mobility, or \emph{churn}, with erasure coding. Communication and storage costs (in units of energy) are considered. We focus on finding the coding method that minimizes the overall cost in the network. Closed-form expressions for the expected energy consumption incurred by data delivery and redundancy maintenance are derived, and it is shown that coding significantly decreases the overall energy consumption -- by more than 90\% in a realistic scenario. It is further shown that D2D caching can also yield notable economical savings for telecommunication operators. Our results are illustrated by numerical examples and verified by extensive computer simulations.
\end{abstract}

\begin{IEEEkeywords}
Device-to-Device Communications, Regenerating Codes, Wireless Caching, Markov Processes, Distributed Data Storage
\end{IEEEkeywords}

\IEEEpeerreviewmaketitle

\section{Introduction}
\label{sec:intro}
\IEEEPARstart{R}{ecent} years have seen an unprecedented growth in wireless data traffic and this growth is not slowing down. Compared to 2016, aggregate smartphone traffic is expected to increase almost tenfold by 2020 \cite{cisco}. One
promising technology to help meet the needs of heavily loaded future cellular networks is \emph{Device-to-Device} (D2D) communications. The major benefit of D2D is that it allows for direct communication between proximate user equipment without the need of base stations, hence potentially offering higher data transfer speeds, lower latency, decreased interference, increased spectral efficiency and lower overall power consumption \cite{janis,doppler,fodor,yu,asadi}.

Another uprising technology is wireless \emph{caching} at either directly on user terminals \cite{maddah, jeon, gerami, guocooperative}, or both user terminals and base stations \cite{altman, golrezaeibase, bastug, jidisse}. Wireless D2D caching is an enticing future technology where data could be stored and distributed directly between mobile terminals -- especially if the involved mobile terminals are geographically close to each other and can thus form D2D \emph{clusters} \cite{afshang}. Geographically constrained caching is of particular interest since the popularity of data is highly location dependent \cite{jidisse}.

Wireless content caching and data distribution through direct links have been proposed in several works such as \cite{ott}, where delay-tolerant networking is considered for message dissemination and forwarding. In \cite{lenders} a wireless peer-to-peer type of application is studied and it is shown that caching can greatly increase the application-level throughput. The potential of coded wireless D2D caching is investigated in \cite{jipaper}, while \cite{golrezaeibase} shows that D2D caching can improve the throughput of wireless video transmission. A method for minimizing the energy consumption of D2D caching nodes is analyzed in \cite{chenenergy}, whereas a joint transmission and caching policy that reduces both the total energy consumption at the base station and the economical cost for the operator is presented in \cite{gregori}. In \cite{afshang}, the authors study clusters-centric D2D networks and demonstrate significant improvements in the network performance.

Joint use of caching and erasure coding for D2D clusters has been proposed in our previous work \cite{globecom,macom} for instantaneous repairs. This work has been extended in \cite{pedersen} to efficiently scheduled repairs. Further work on distributed storage with D2D communications has been done in \cite{wangwu}, where a combination of D2D and social networks is considered. In \cite{globecom} we looked for a way to strictly minimize the amount of data traffic in caching clusters and found that repetition coding yields the best results for the considered system model. We then found in \cite{macom} that the optimal coding method, \emph{i.e.}, the coding method which minimized a predetermined cost function, highly depends on the popularity of the file.

A clear drawback of geographically constrained wireless caching is unconstrained user mobility -- when a caching node moves away from the caching cluster, its content is lost. To avoid this, we introduce erasure coding to ensure data availability. The focus of this article is studying the performance of such coded caching clusters. The main contributions of this article can be summarized as follows:
\begin{itemize}
  \item We construct a system model for a clustered wireless D2D caching community based on stochastic geometry.
  \item Closed-form expressions for the expected energy cost based on signal attenuation of both uncoded and coded D2D caching methods are derived, and we further examine under which conditions coded caching outperforms uncoded caching without redundancy.
  \item It is shown that coded caching can yield significant cost savings in terms of both the overall energy consumption and economical cost savings from a operator's point of view.
\end{itemize}

The rest of this paper is organized as follows. In Section~\ref{sec:d2d}, we present the system model used throughout this work. In Section~\ref{sec:methods}, we introduce the proposed caching methods. Analytical cost estimates are derived in section Section~\ref{sec:cost}, while simulation results are presented, and compared with the analytical results, in Section~\ref{sec:analysis}. Finally, conclusions are drawn in Section~\ref{sec:conclu}.

\section{System Model}
\label{sec:d2d}

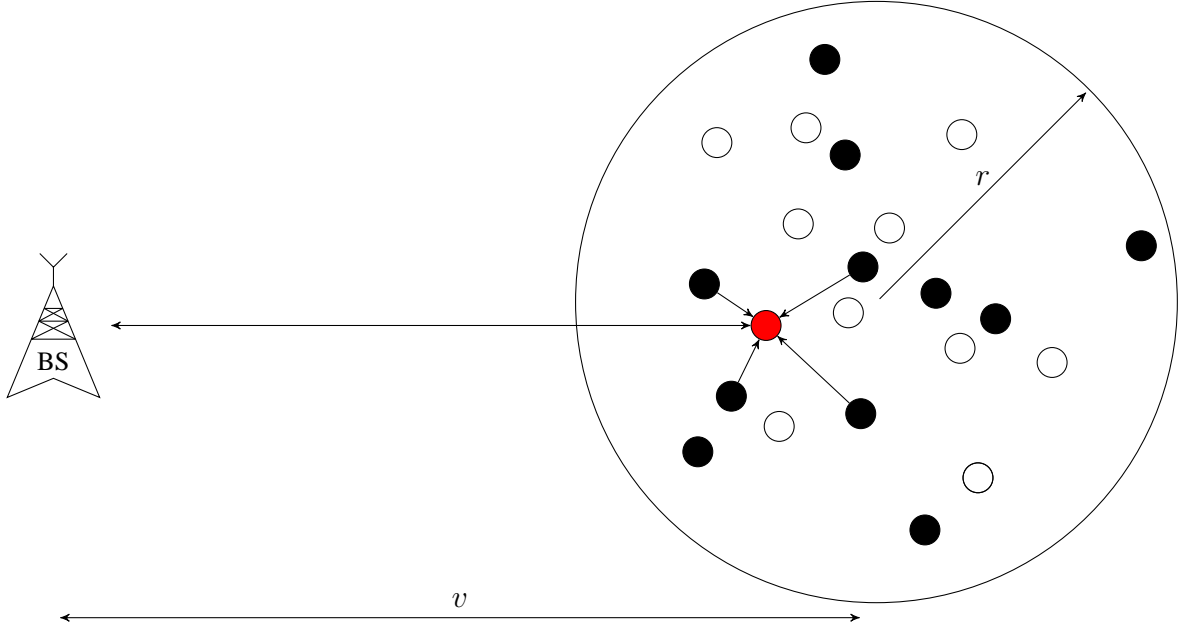
\begin{figure}[!h]
\centering
\begin{tikzpicture}
\draw (0,0) circle (4cm);
\node[draw,shape=circle,fill=black,minimum size=.01cm] (v0) at (111:.5) {$ $};
\node[draw,shape=circle,fill=black,minimum size=.01cm] (v1) at (8.4:.8) {$ $};
\node[draw,shape=circle,fill=red,minimum size=.01cm] (v3) at (192:1.5) {$ $};
\node[draw,shape=circle,fill=black,minimum size=.01cm] (v4) at (213:2.3) {$ $};
\node[draw,shape=circle,fill=black,minimum size=.01cm] (v5) at (262:1.5) {$ $};
\node[draw,shape=circle,fill=black,minimum size=.01cm] (v6) at (282:3.1) {$ $};
\node[draw,shape=circle,fill=black,minimum size=.01cm] (v7) at (102:3.3) {$ $};
\node[draw,shape=circle,fill=black,minimum size=.01cm] (v8) at (12:3.6) {$ $};
\node[draw,shape=circle,fill=black,minimum size=.01cm] (v9) at (352:1.6) {$ $};
\node[draw,shape=circle,fill=black,minimum size=.01cm] (v10) at (220:3.1) {$ $};
\node[draw,shape=circle,fill=black,minimum size=.01cm] (v11) at (174:2.3) {$ $};
\node[draw,shape=circle,fill=black,minimum size=.01cm] (v12) at (102:2.0) {$ $};
\node[draw,shape=circle,minimum size=.01cm] (v13) at (201:.4) {$ $};
\node[draw,shape=circle,minimum size=.01cm] (v14) at (112:2.5) {$ $};
\node[draw,shape=circle,minimum size=.01cm] (v15) at (232:2.1) {$ $};
\node[draw,shape=circle,minimum size=.01cm] (v16) at (63:2.5) {$ $};
\node[draw,shape=circle,minimum size=.01cm] (v17) at (300:2.7) {$ $};
\node[draw,shape=circle,minimum size=.01cm] (v18) at (300:2.7) {$ $};
\node[draw,shape=circle,minimum size=.01cm] (v19) at (80:1) {$ $};
\node[draw,shape=circle,minimum size=.01cm] (v20) at (135:1.47) {$ $};
\node[draw,shape=circle,minimum size=.01cm] (v21) at (331:1.27) {$ $};
\node[draw,shape=circle,minimum size=.01cm] (v21) at (341:2.47) {$ $};
\node[draw,shape=circle,minimum size=.01cm] (v21) at (135:3.00) {$ $};
\node[left=8.5cm of v3] (BS) {\BS{BS}};
\node[draw=none,minimum size=.00cm] (cen) at (225:.15) {$ $};
\node[draw=none,minimum size=0cm] (bor) at (45:4.15) {$ $};
\node[draw=none,minimum size=.00cm] (BS1) at (200.8978:11.7745) {};
\node[draw=none,minimum size=.00cm] (BS2) at (269:4.2) {};

\draw[->] (v11) --  node[above] {} (v3);
\draw[->] (v0) --   node[above] {} (v3);
\draw[->] (v4) --   node[left] {$ $} (v3);
\draw[->] (v5) --   node[above] {$ $} (v3);
\draw[<->] (BS) --  node[above] {$ $} (v3);
\draw[->] (cen) --  node[above] {$r$} (bor);
\draw[<->] (BS1) -- node[above] {$v$} (BS2);
\end{tikzpicture}
\caption{D2D caching cluster system model. Instead of contacting a remote base station, users in the cluster are able to communicate with each other through direct links.}
\label{fig:cluster}
\end{figure}

We begin by introducing the system model assumed throughout the paper. We model a cluster of mobile terminals with data storage capabilities -- or \emph{nodes} -- by a disk of radius $r$. The expected number of nodes present in the cluster is denoted by $m$, and the nodes are assumed to be uniformly distributed inside the disk. A single base station is located at a distance $v>r$ from the center of the cluster. A graphical representation of the model is displayed in Figure~\ref{fig:cluster}.

The nodes inside the cluster form a D2D caching community. We assume that each node knows about the content stored in every other node, and any two nodes can communicate data. We further assume that all data transmission links are error-free. 

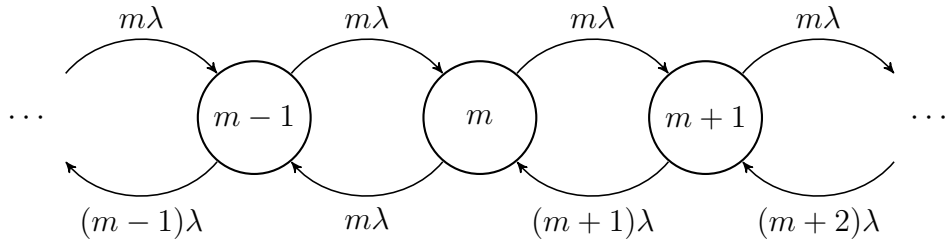
\begin{figure}[!h]
\centering
\begin{tikzpicture}[->, >=stealth', auto, semithick, node distance=3cm]
\tikzstyle{every state}=[fill=white,draw=black,thick,text=black, minimum width=1.5cm]
\node[state] (A)               {$m-1$};
\node[state] (B) [right of=A]  {$m$};
\node[state] (C) [right of=B]  {$m+1$};
\node[state,draw=none] (d1)[left of=A]  {$\cdots$};
\node[state,draw=none] (d2)[right of=C] {$\cdots$};
\path (A)  edge[bend left=50,above] node{$m\lambda$}      (B)
		   edge[bend left=50,below] node{$(m-1)\lambda$}  (d1)
	  (B)  edge[bend left=50,above] node{$m\lambda$}      (C)
           edge[bend left=50,below] node{$m\lambda$}      (A)
      (C)  edge[bend left=50,above] node{$m\lambda$}		 (d2)
	       edge[bend left=50,below] node{$(m+1)\lambda$}  (B)
      (d1) edge[bend left=50,above] node{$m\lambda$} 	 (A)
      (d2) edge[bend left=50,below] node{$(m+2)\lambda$}  (C);
\end{tikzpicture}
\caption{M/M/$\infty$ Markov chain. The state refers to the number of users in the cluster.}
\label{fig:markov_system}
\end{figure}

The time dynamics of the system are modeled as follows. The time that an arbitrary node remains active in the cluster follows an exponential distribution with expected value $T$. We define a \emph{failure} as the event when a node becomes inactive by leaving the system and denote the \emph{node failure rate} by $\lambda = 1/T$. With these parameters, we can model the instantaneous state of the system via an M/M/$\infty$ Markov model (cf. Figure~\ref{fig:markov_system}), which has been widely used to model wireless cellular systems with exponential dwell times \cite{tanganalysis, hungrandom, thaj}. In this work, we only consider the steady state of the chain with $m$ nodes in the cluster on average. Hence, the probability that the system is in state $j$, \emph{i.e.}, that there are $j$ nodes in the cluster, can be written as \cite{harrison}
\begin{align}
	\pi(j) = \frac{m^j}{j!}e^{-m}.\label{marko}
\end{align}

We henceforth consider a single data file of unit size without loss of generality. Each user in the cluster can request the file anytime. The request interval of a user follows an exponential distribution with expected value $1/\omega$, where we call $\omega$ the \emph{request rate} or, by slight abuse of terminology, the \emph{file popularity}. We concentrate on the case $\omega < \lambda$ as we assume that the vast majority of the users request the file only once during their visit to the cluster.

\section{Caching Methods}
\label{sec:methods}

In this article, we consider three different methods to cache the file on the nodes. These caching methods are introduced in the following subsections.

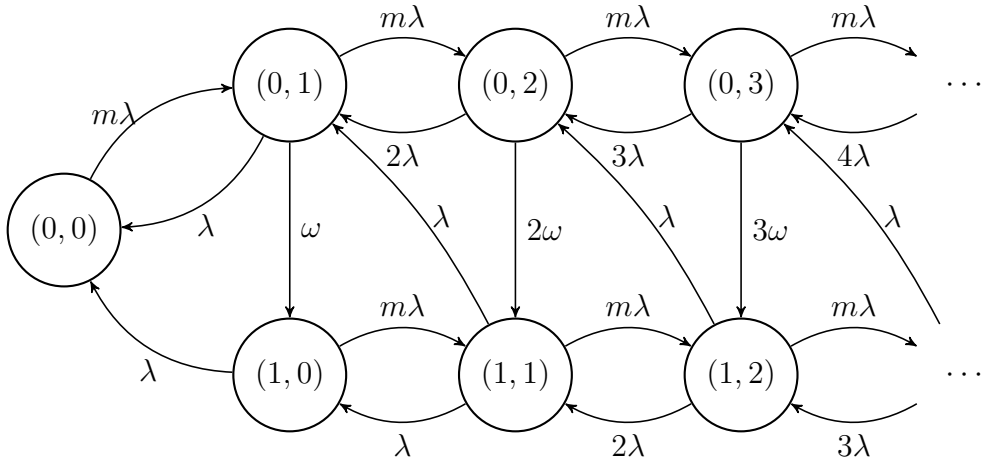
\begin{figure}[!h]
\centering
\begin{tikzpicture}[->, >=stealth', auto, semithick, node distance=3cm]
\tikzstyle{every state}=[fill=white,draw=black,thick,text=black, minimum width=1.5cm]
\node[state]    (A)              			   			{$(0,0)$};
\node[state]    (B)	 [right of=A, below=0.4cm of A]		{$(1,0)$};
\node[state]    (C)	 [right of=A, above=0.4cm of A]  	{$(0,1)$};
\node[state]    (D)	 [right of=B]  						{$(1,1)$};
\node[state]    (E)	 [right of=C]  						{$(0,2)$};
\node[state]		(d2) [right of=D] 						{$(1,2)$};
\node[state]		(d3) [right of=E] 						{$(0,3)$};
\node[state,draw=none] (d4)[right of=d3] 				{$\cdots$};
\node[state,draw=none] (d5)[right of=d2] 				{$\cdots$};

\path
(A)  edge[bend left=30,left]      node{$m\lambda$}   (C)
(B)  edge[bend left=30,above]     node{$m\lambda$}   (D)
	 edge[bend left=30,below]     node{$\lambda$}    (A)
(C)  edge[bend left=30,above]     node{$m\lambda$}   (E)
	 edge[bend left=30,below]     node{$\lambda$}    (A)
	 edge[bend left=0,right]      node{$\omega$}     (B)
(D)  edge[bend left=30,above]     node{$m\lambda$}   (d2)
	 edge[bend left=-10,right]    node{$\lambda$}    (C)
	 edge[bend left=30,below]     node{$\lambda$}    (B)
(E)  edge[bend left=30,above]     node{$m\lambda$}   (d3)
	 edge[bend left=30,below]     node{$2\lambda$}   (C)
	 edge[bend left=0,right]      node{$2\omega$}    (D)
(d2)
	 edge[bend left=30,above]     node{$m\lambda$}   (d5)
	 edge[bend left=30,below]     node{$2\lambda$}   (D)
	 edge[bend left=-10,right]    node{$\lambda$}    (E)
(d3)
	 edge[bend left=30,below]     node{$3\lambda$}   (E)
	 edge[bend left=0,right]      node{$3\omega$}    (d2)
	 edge[bend left=30,above]     node{$m\lambda$}   (d4)
(d4)
	 edge[bend left=30,below]     node{$4\lambda$}   (d3)
(d5)	
	 edge[bend left=30,below]     node{$3\lambda$}   (d2)
	 edge[bend left=-10,right]    node{$\lambda$}    (d3);
\end{tikzpicture}
\caption{Simple caching Markov chain state diagram. State $(x,y)$ refers to having $x\in\{0,1\}$ caching nodes and $y=0,1,2,3,...$ empty nodes in the cluster.}
\label{fig:markov_caching}
\end{figure}

\paragraph{Simple caching} A single node stores a full copy of the file. The file is not protected against storage node failures since no redundancy is enabled. As soon as the caching node leaves the system, the data file is lost from the caching community and the next requesting node needs to download the entire file from the base station. This node then automatically becomes the new caching node and, as long as it remains active in the cluster, all file requests from other nodes are served by this node through D2D communications. The system can be modeled with a Markov chain as depicted in Figure~\ref{fig:markov_caching}.

The steady state probabilities of the upper chain are $\pi_j-\zeta_j$ and the lower chain $\zeta_j$, where $\pi_j$ are the M/M/$\infty$ probabilities from \eqref{marko}, and $\zeta_j$ fulfil the recursion $$
\zeta_{j+1} = \left(\frac{m}{j} + \frac\omega\lambda + 1\right)\zeta_j - \frac{m}{j}\zeta_{j-1} - \frac\omega\lambda \pi_j~,
$$
with $\zeta_0=0$. Note that, for the purposes of this article, we do not need to find the steady state probabilities. Instead, in Section \ref{subsec:savings_general}, we derive an approximation of the performance metric. We use the chain of Figure \ref{fig:markov_caching} only to model the behavior of the system with computer simulations in order to empirically measure the performance of simple caching. This will be done later in Section \ref{sec:analysis}.

\begin{remark}
Another way to cache and disseminate the file in the cluster would be to store a replica of the file on each of the nodes that requests it. However, it is easy to see that in order for this method to work and the cluster to fill up with replicas, the file request rate should be higher than the node passing rate, i.e., $\omega > \lambda$. This in turn would mean that the average user downloads the file more than once during its stay in the cluster. We focus on the more realistic case where the average number of requests per node lifetime is less than one, i.e., $\omega < \lambda$, which means that redundancy must be actively maintained or else the cached data will be lost.
\end{remark}

\paragraph{Replication}
The most elementary way of adding redundancy to the system is simply to store multiple copies of the entire file on separate nodes. We refer to this strategy as $n$-replication, where $n\ll m$ nodes store a replica of the file. When the system operates under this method, the file can be retrieved, or a lost node repaired, by contacting simply one of the storage nodes. The obvious downside of replication is that it consumes more storage space than coded storage. Furthermore, the \emph{repair bandwidth}, that is, the amount of data traffic that replacing a lost storage node incurs, is equal to the size of the entire file. Hence, the repair bandwidth is equal to the \emph{reconstruction bandwidth}, which we define as the amount of data traffic incurred when a users downloads and reconstructs the data file.

\paragraph{Regenerating Codes} 
We interpret the considered system as a \emph{Distributed Storage System} (DSS) which is composed of $n \ll m$ storage nodes\footnote{With a slight abuse of notation, we denote by $n$ the number of nodes storing a replica in case of $n$-replication, and the length of an $(n,k,d)$ MDS code used for the DSS. The meaning of $n$ will always be clear from the context or clarified otherwise.}. The original data file is encoded into $n$ coded fragments of size $\alpha$ each. Storage nodes are assigned one of the coded fragments, and the entire file can be recovered by contacting any $k < n$ storage nodes, a feature also referred to as the \emph{Maximum Distance Seperability} (MDS) property of a code. This property is what allows the system to be resistant against arbitrary failure sequences.

To maintain redundancy, whenever a storage node fails, it is instantly replaced with a \emph{newcomer} node that is randomly chosen from the empty nodes present in the cluster. This newcomer node contacts any $d \le n-1$ storage nodes, downloads $\beta$ units of data from each and stores $\alpha$ units of data. Note that the new content in the newcomer node does not need to be exactly the data that were lost in the failed node. Hence, we consider \emph{functional repair}, which ensures that both the MDS and the regeneration property hold after an arbitrary failure.

Throughout this paper, we assume instant repair after failures so that no matter which coding method is used, there are always $n$ caching nodes in the cluster as long as the Markov chain in Figure \ref{fig:markov_system} never goes to a state lower than $n$, which we deem a valid assumption as we only investigate the case $n \ll m$, and thus the probability of finding the chain in small states is extremely small\footnote{For example, if $m=100$ and $n=6$, values which we will later use in our simulations, the probability that the number of nodes in the cluster drops to $n$ or below is approximately $5.5 \times 10^{-35}$.}.

A DSS is determined by the tuple $(n,k,d,\alpha,\gamma)$, whereof the triple $(n,k,d)$ consists of the \emph{storage degree}, \emph{reconstruction degree} and \emph{repair degree}. In other words, reconstructing the data file requires contacting $k$ out of total $n$ storage nodes, while repairing the contents of a lost node requires contacting $d$ nodes. In addition, the parameter tuple $(\alpha,\gamma)$ consists of the fragment size $\alpha$ stored in each of the $n$ storage nodes, and the \emph{repair bandwidth} $\gamma$, that is the total number of units of data that a newcomer needs to download for repairing a lost node. Note that when repairing, each storage node involved in the repair process transmits $\beta$ units of data to the newcomer node, so that $\gamma = d\beta$.

A given tuple of parameters $(n,k,d,\alpha,\gamma)$ is \emph{feasible} if a code with such $\alpha$ and $\gamma$ exists. For a result on the existence of feasible parameter tuples, we refer to \cite[Thm.~1]{dimakis}. More importantly, there is a natural tradeoff between $\alpha$ and $\gamma$ given by a piecewise linear function. Codes lying on this tradeoff curve are called regenerating codes. Hence, regenerating codes offer an optimal tradeoff between storage space consumption and repair bandwidth, while maintaining the MDS property. Furthermore, any $d$ nodes can be contacted to resurrect a lost node while maintaining these properties after repairs. Hence, regenerating codes are an attractive choice.

In this work, we consider two types of regenerating codes: codes attaining one of the two extremal points, \emph{i.e.}, the points where either the storage space consumption or repair bandwidth is minimized. These codes are known as \emph{minimum storage regenerating} (MSR) codes and \emph{minimum bandwidth regenerating} (MBR) codes, respectively. For a file of unit size, these points are achieved by the pairs \cite{dimakis}
\begin{align}
	\left(\alpha_{\text{MSR}},\gamma_{\text{MSR}}\right) &= \left(\frac{1}{k},\frac{d}{k(d-k+1)}\right), \label{MSRab}
\\
	\left(\alpha_{\text{MBR}},\gamma_{\text{MBR}}\right) &= \left(\frac{2d}{k(2d-k+1)},\frac{2d}{k(2d-k+1)}\right). \label{MBRab}
\end{align}
It has been shown that, in the typical case $k \leq d \leq n-1$ which we assume throughout this work, code constructions exists for both the MSR and the MBR point, see \emph{e.g.}, \cite{rashmi}. Note that the reason we do not consider traditional MDS erasure codes, such as Reed-Solomon codes, is that, for the purpose of this work, they are merely a special case of MSR codes with $k = d$.

\section{Cost Estimates}
\label{sec:cost}

In order to compare the three considered methods, we need to determine a reference function which measures the overall expected costs in terms of transmission energy. We start by establishing a general underlying model. 

The main performance metric of the system is the overall energy \emph{cost}, which we define as the sum of the \emph{transmission cost} and the \emph{storage cost}. We refer to the {transmission cost} of a transmission scheme as the sum of the expected overall transmission costs, that is, the transmit power consumption of the base station and D2D community caused by data traffic of a fixed file of unit size, both due to data retrieval or repair. In addition, we also establish a {storage cost}, so that neglectfully caching large amounts of data is not a viable option. Wasting storage space would result in a waste of transmission energy as the short-distance D2D links could not be efficiently utilized if only a few different files fit on the storage space of the caching community, and consequently, the traditional downlink with the base station would be needed more often. Hence, we \emph{translate storage into transmit power}.

We represent the cost of storing a unit of data by a constant $\sigma$. Finding the data transmission costs requires analyzing the stochastic geometrical properties of the cluster, which we will do in the following to derive the cost of reconstruction and repair.

As depicted in Figure~\ref{fig:cluster}, our system consists of a base station located at a distance $v$ away from the center of the caching cluster, and a cluster of nodes, uniformly distributed in a disk of radius $r \ll v$. We implement full channel inversion at the transmitter, which implies that the expected required downlink transmit power to communicate one unit of data over distance $x$ becomes $x^{\Gamma}$, where $\Gamma$ is the pathloss exponent. We consider two different pathloss exponents\footnote{We assume that the base station antennas are located much higher than the D2D users. Without loss of generality, throughout this work we use values $\Gamma_\text{BS} = 2$ for the pathloss between the base station and a node in the cluster, and $\Gamma_\text{D2D} = 4$ for the pathloss between two nodes in the D2D community, when numerical values are needed similarly to, \emph{e.g.}, \cite{guodevice}.}: one for the downlink from the base station to the nodes in the cluster ($\GB$), and another for communications in D2D mode ($\GD$).

The expected required transmit power for communication between two nodes in the cluster is denoted by $L_{r,\GD}(i,n)$, which is the expected $\GD^\text{th}$ power of the distance from an arbitrary node in the disk to its $i^\text{th}$ nearest caching node, assuming that there are $n$ uniformly distributed storage nodes present in a disk of radius $r$. In other words, $L_{r,\GD}(i,n)$ is also the expected cost of transmitting a unit of data between two nodes in the disk. Thus, the first step towards estimating the transmission costs of the individual methods is to derive the quantity $L_{r,\Gamma}(i,n)$. To that end, we will need the following result.

	\emph{Let two circles of radii $R$ and $r\leq R$ be separated by distance $v$. For any triple $(R,r,v)$, the intersection area $A(R,r,v)$ of the two circles is given by the function }
	\begin{align}
		A(R,r,v) = \begin{cases}
			\pi r^2 &\mbox{ if } v \le R-r \\
			\pi r^2 - \eta(r,\mu) + \eta(R,\mu) &\mbox{ if } R-r < v \le \sqrt{R^2-r^2} \\
			\eta(r,\mu) + \eta(R,\mu) &\mbox{ if } \sqrt{R^2-r^2} < v \le r+R \\
			0 &\mbox{ if } v > r+R,
		\end{cases}
	\end{align}
	where
	\begin{align}
		\mu \coloneqq \mu(R,r,v) &\coloneqq \frac{1}{v}\sqrt{(r+R-v) (r-R+v) (-r+R+v) (r+R+v)},\nonumber \\
		\eta(x,\mu) &\coloneqq x^2\sin^{-1}\left(\frac{\mu}{2x}\right) - \sqrt{\left(\frac{\mu}{2}+x\right) \left(\left(\frac{\mu}{2}+x\right)-\mu\right) \left(\left(\frac{\mu}{2}+x\right)-x\right)^2}.\nonumber
	\end{align}

For further details on circle intersection calculations, see \emph{e.g.} \cite{circlecircle}. Now let $P(t)$ be a node in the cluster, where $t$ denotes its distance from the origin of the disk. Using the computed area of intersection, we can find the probabilities needed for our calculations. Of interest for our purposes is the expected distance between the node $P(t)$ and it's $q^\text{th}$ nearest node out of $n$ nodes, which can be computed as (see \cite{srinidist} for further details)
\begin{align}
	E(n,q,r,t) = &\int\limits_{0}^{r}\left(\sum\limits_{i = 0}^{q-1}{\binom{n}{i}}\left(\frac{A(r,x,t)}{\pi r^2}\right)^i\left(1-\frac{A(r,x,t)}{\pi r^2}\right)^{n-i}\right)dx \nonumber \\
	+ &\int\limits_{r}^{r+t}\left(\sum\limits_{i = 0}^{q-1}{\binom{n}{i}}\left(\frac{A(x,r,t)}{\pi r^2}\right)^i\left(1-\frac{A(x,r,t)}{\pi r^2}\right)^{n-i}\right)dx.\nonumber
\end{align}

Moreover, we are interested in the expected value of the $\Gamma^\text{th}$ power of the distance between $P(t)$ and its $q^\text{th}$ nearest neighbor, which becomes 
\begin{align}
	\mathcal{E}_{\Gamma}(n,q,r,t) =  \Gamma&\left(\int\limits_{0}^{r}x^{\Gamma-1}\left(\sum\limits_{i = 0}^{q-1}{\binom{n}{i}}\left(\frac{A(r,x,t)}{\pi r^2}\right)^i\left(1-\frac{A(r,x,t)}{\pi r^2}\right)^{n-i}\right)dx\right. \nonumber \\
	+ &\left.\int\limits_{r}^{r+t}x^{\Gamma-1}\left(\sum\limits_{i = 0}^{q-1}{\binom{n}{i}}\left(\frac{A(x,r,t)}{\pi r^2}\right)^i\left(1-\frac{A(x,r,t)}{\pi r^2}\right)^{n-i}\right)dx\right),\nonumber
\end{align}
the expected value of which is given by
\begin{align}
	L_{r,\Gamma}(q,n) = \frac{2}{r^2}\int\limits_{0}^{r}t\mathcal{E}_{\Gamma}(n,q,r,t)dt,
\end{align}
where we have used the probability density function $f(t)=\frac{2t}{r^2}\quad (0\leq t\leq r)$ corresponding to the random variable representing the distance between a randomly chosen point in a disk of radius $r$ and the center of the disk.
Lastly, to measure the performance of simple caching, we find the expectation of the $\GB^\text{th}$ power of the distance from a node in the cluster to the base station by integrating the complementary cumulative density function of the distance:
\begin{align}
	\mathcal{E}_{\GB}(r,v) = \GB \int\limits_{0}^{v+r}{x^{\GB-1}\left(1-\frac{A(x,r,v)}{\pi r^2}\right) dx}.
	\label{EBS}
\end{align}
For clarity, the notation is summarized in the appendix in Table~\ref{tab:parameters}.

\subsection{Cost Functions Considering Overall Energy Savings}
\label{subsec:savings_general}

We begin by finding the costs of each of the considered caching methods. We only consider the expectations of the costs and thus directly use the expected numbers of nodes to perform calculations. We later verify the validity of this approach with computer simulations in Section~\ref{sec:analysis}.

\begin{enumerate}
	\item \underline{Simple Caching.} The dynamics of the system under simple caching are modeled according to the Markov chain in Figure~\ref{fig:markov_caching}. Instead of a full steady state analysis of the chain, for the sake of simplicity, we derive an approximation for the expected cost in the following. When the file is cached, there is one node caching the entire file with no redundancy, so the cost of repair vanishes. There are, on average, $m-1$ nodes in the cluster generating requests as the single caching node does not need to download the file itself. Thus, the expected number of requests during the lifetime of the caching node is $(m-1)\omega T = (m-1)\frac{\omega}{\lambda}$. Once the caching node leaves the cluster, the next file request will be directed to the base station. The expected time in which this happens is approximately\footnote{Strictly speaking, when the caching node has left the cluster, we should take the transient period in which the system returns back to steady state into account to find the exact expected value of the cost of simple caching. Since $\lambda < \omega$ and $m$ is large, though, this approximation is accurate enough for our purposes as will be demostrated later by the numerical results.} $\frac{1}{m\omega}$, and an expected number of $(m-1)\frac\omega\lambda + 1$ requests, including the local file retrievals in the cluster and the remote retrieval from the base station, are generated in time $T+\frac{1}{m\omega}=\frac 1\lambda+\frac{1}{m\omega}$.
	The cost of retrieving the file from the caching node is $L_{r,\GD}(1,1)$, whereas the cost of retrieving it from the base station is $\mathcal{E}_{\GB}(r,v)$. Further, as long as the file is cached, it incurs a storage cost of $\sigma$. Using the approximation\footnote{This follows from the Taylor series expansion of $f(x,y)=x/y$ centered at the point $\left(E(X),E(Y)\right)$ when $y$ has support on $[0,\infty)$. This expansion can be truncated to $E(X/Y) \approx E(X)/E(Y) - \text{Cov}(X,Y)/E(Y)^2 + \text{Var}(Y)E(X)/E(Y)^3$ \cite{kendall}. In the interest of space, instead of providing a full analysis of the error term, we will demonstrate the predictive ability of our estimate through numerical simulations, see Figures \ref{fig:sim}, \ref{fig:simMSRrocks} and \ref{fig:reprules}.} $E(X/Y) \approx E(X)/E(Y)$, where $X,Y$ are two random variables and $E(\cdot)$ denotes expectation, the cost of simple caching can be approximated as 
	\begin{align}
			\chi{\scriptstyle{(\text{Simple Caching})}} \approx  \frac{(m-1)\frac{\omega}{\lambda} L_{r,\GD}(1,1) + \mathcal{E}_{\GB}(r,v) +  \sigma}{\frac{1}{\lambda}+\frac{1}{m\omega}}.
	\label{scequ}
	\end{align}

The accuracy of this approximation, in the special cases considered in this work, is verified by numerical results in Section~\ref{sec:analysis}. Note that there is nothing we can optimize about this caching method -- we use \eqref{scequ} only as a baseline to measure the improvement achieved by storage coding methods ``replication" and ``regenerating codes" which will be introduced in the following.
	
	\item \underline{Replication.} When replication is used, we assume $n$ storage nodes storing an entire replica of the file. On average, there are $m-n$ empty nodes each of which generates file requests at rate $\omega$. For reconstructing the file, the requesting node contacts the nearest storage node, so that the reconstruction cost is $L_{r,\GD}(1,n)$, so the reconstruction cost becomes
	\begin{align*}
	(m-n)\omega L_{r,\GD}(1,n).
	\end{align*}
To repair a failed node, the newcomer node contacts the nearest out of the surviving $n-1$ storage nodes. The repair cost is hence given by $L_{r,\GD}(1,n-1)$. Thus, as there are $n$ storage node each failing at rate $\lambda$, the reconstruction cost becomes
	\begin{align*}
	n\lambda L_{r,\GD}(1,n-1).
	\end{align*}
The storage cost in this scenario is simply $n\sigma$. Now recall that each node in the cluster generates requests at rate $\omega$, and each node passes through the cluster at rate $\lambda$. Therefore, the cost of replication becomes
	\begin{align}
		\chi{\scriptstyle{(\text{Replication})}} = (m-n)\omega L_{r,\GD}(1,n) + n\lambda L_{r,\GD}(1,n-1) + n\sigma.
	\label{repequ}
	\end{align}
The only parameter to be optimized for replication is the number of replicas $n$. Examining \eqref{repequ}, it is a straightforward, yet important observation that increasing $n$ decreases the expected distances between the nodes and the number of empty nodes that request the file, but increases the total failure rate, and consequently the total repair cost, and the total storage cost. Note that similar observations have been made before for similar distance-dependent cost functions, see \emph{e.g.} \cite{altman} and references therein. For the purposes of our work we emphasize that to minimize the cost of replication it is crucial to find a suitable value of $n$, as will be demonstrated later in this work.
	
	\item \underline{Regenerating Codes.} In a system operating under this scheme, there are both storage nodes storing a fragment of the data file and empty nodes present in the cluster. We hence need to consider two types of requests. When one of the $n$ storage nodes requests the file, it contacts $k-1$ out of the remaining $n-1$ storage nodes and downloads $\alpha$ units of data from each, which yields cost
\begin{align*}
n\omega \alpha\sum\limits_{i=1}^{k-1}{L_{r,\GD}}(i,n-1).
\end{align*}
When one of the empty nodes requests the file, $k$ out of the $n$ storage nodes need to be contacted, thus yielding a cost
\begin{align*}
(m-n)\omega \alpha\sum\limits_{i=1}^{k}{L_{r,\GD}(i,n)}
\end{align*}
since the expected number of empty nodes in the cluster is $m-n$.
	
When a storage node is lost, one of the empty nodes acts as the newcomer, contacts $d$ of the remaining $n-1$ surviving nodes, and downloads $\beta$ units of data from each, generating a total repair bandwidth of $\gamma = d\beta$. Thereby, the repair cost becomes
\begin{align*}
n\lambda \beta\sum\limits_{i=1}^{d}{L_{r,\GD}(i,n-1)}.
\end{align*}
The storage cost using regenerating codes is simply $n\alpha\sigma$, so the total cost of using regenerating codes amounts to 
	\begin{empheq}[box=]{align}
			\chi{\scriptstyle{(\text{Regenerating})}} =\ &n\omega \alpha\sum\limits_{i=1}^{k-1}{L_{r,\GD}}(i,n-1) +(m-n)\omega \alpha\sum\limits_{i=1}^{k}{L_{r,\GD}(i,n)} \nonumber \\
			&+ n\lambda \beta\sum\limits_{i=1}^{d}{L_{r,\GD}(i,n-1)} + n\alpha\sigma,
	\label{regeequ}
	\end{empheq}
\end{enumerate}
where $\alpha$ and $\beta$ are functions of $(k,d)$ and are given by \eqref{MSRab} for MSR and \eqref{MBRab} for MBR codes.

We immediately see that the same observations about varying the storage degree $n$ that we made for replication apply to \eqref{regeequ} as well. Further, for regenerating codes we also need to choose the optimal values of $k$ and $d$, as well as either the MSR or MBR point, to minimize the cost for given system parameters and file popularity. Maximizing the repair degree $d$ minimizes both $\alpha$ and $\beta$ for MBR and $\beta$ for MSR, and maximizing the reconstruction degree $k$ minimizes the amount of redundancy for MSR. However, high values of $k$ and $d$ imply that distant nodes need to be contacted, and as the transmission cost is proportional to the $\Gamma^\text{th}$ power of the distance, we conclude that naively ignoring the distance-dependency and only optimizing with regard to the amount of data traffic does not necessarily imply the lowest cost.

\subsection{Savings from an Operator's Point of View}
\label{subsec:operator}

So far we have been only concerned with saving overall transmission power by taking advantage of both caching on devices and direct data transmission between users. However, users in the cluster can be selfish in nature and thus may not have a motive for sharing their storage and battery to enable a caching system such as the one presented in this work. Thus, we now focus on the case where we assume that users sharing their resources are rewarded by the operator with lower charges if maintaining the community implies economical savings for the operator. Similar incentives have been proposed earlier in the literature \cite{chencaching,wugame,alogame,gregori}.

It is a natural question to ask whether from an operator's point of view the maintenance -- or \emph{upkeep} -- of such a D2D community pays off. To measure economical profit, we consider the ratio of the costs of downlink transmissions, and D2D traffic and upkeep costs.

Deriving the cost of traditional downlink communications is straightforward. There are $m$ nodes generating requests at frequency $\omega$, and this cost thus amounts to 
\begin{align*}
	\chi{\scriptstyle{(\text{Downlink})}} = m\omega \mathcal{E}_{\GB}(r,v),
\end{align*}
where $\mathcal{E}_{\GB}(r,v)$ is as in \eqref{EBS}.

To weigh the costs of D2D data transmission and storage, we say that transmitting a unit of data in D2D mode over unit distance incurs a cost $\Theta$ for the operator, while storing a unit of data costs $\sigma$. In other words, these are the incentives offered to a caching user: $\Theta$ represents the economical benefit that a caching user gains from distributing data by using transmit power and $\sigma$ represents the benefit that a caching user gains when storing data. Now similarly to \eqref{scequ}, \eqref{repequ} and \eqref{regeequ}, we find the upkeep costs to be
\begin{align}
	\chi{\scriptstyle{(\text{Upkeep})}} = \begin{cases}
		\Theta\cdot \chi{\scriptstyle{(\text{Simple Caching})}} - \frac{\sigma(\Theta-1)}{\frac{1}{\lambda}+\frac{1}{m\omega}} &\mbox{ for simple caching}, \\
		\Theta \cdot\chi{\scriptstyle{(\text{Replication})}} - n\sigma(\Theta-1) &\mbox{ for replication}, \\
		\Theta\cdot \chi{\scriptstyle{(\text{Regenerating})}} - n\alpha\sigma(\Theta-1) &\mbox{ for regenerating codes}.
	\end{cases}\nonumber
\end{align}
In other words, when data are distributed and cached redundantly, the operator avoids the cost of data transmission from the base station altogether, but has to pay a cost of $\Theta$ for each transmitted unit of data over unit distance and $\sigma$ for each unit of data cached on a user equipment.

We now define the caching gain of an operator as
\begin{align}
	G = \frac{\chi{\scriptstyle{(\text{Downlink})}}}{\chi{\scriptstyle{(\text{Upkeep})}}},
		\label{opegain}
\end{align}
which we call \emph{operator gain}. In the next section, we present numerical results of both operator gains and overall energy consumption cost savings.

\section{Numerical Results}
\label{sec:analysis}

In this section, we illustrate the performance of the four considered caching methods with respect to the derived performance metric with the help of numerical results. We investigate three cases: low, moderate and high storage cost, while the parameters of replication ($n$) and regenerating codes $(n,k,d)$ are chosen from a small interval so that the cost function is minimized. Further, we study the operator gains for short and long distances from the cluster to the base station. 

For all cases in this section, we fix $m=100$, $\lambda = 1$, $r=1$, $\Gamma_{\text{BS}}=2$ and $\Gamma_{\text{D2D}}=4$, while $\sigma$ is varied. For the overall energy consumption results we fix $v=20$, while two values, $v=10$ and $v=20$, are considered for the operator gain. We choose $n \in [2,6]$ for replication, and $n \in [3,6]$ for regenerating codes, so that the cost is minimized for a given $\omega$. The theoretical curves (solid lines) in the figures are numerical values using the derived cost functions \eqref{scequ},  \eqref{repequ} and \eqref{regeequ}, while the simulated values (dots) are obtained by computing steady state averages of long Monte Carlo simulations for the Markov chains depicted in Figure~\ref{fig:markov_caching} for simple caching and Figure~\ref{fig:markov_system} for regenerating codes and replication to verify the theoretical calculations. For all simulations, the initial number of nodes in the cluster is $m=100$ and the file is cached when the simulation starts. The simulation length is $10^4$ expected node lifetimes $T=1$ for each data point.

\begin{figure}[!h]
\centering
	\includegraphics[width=.7\textwidth]{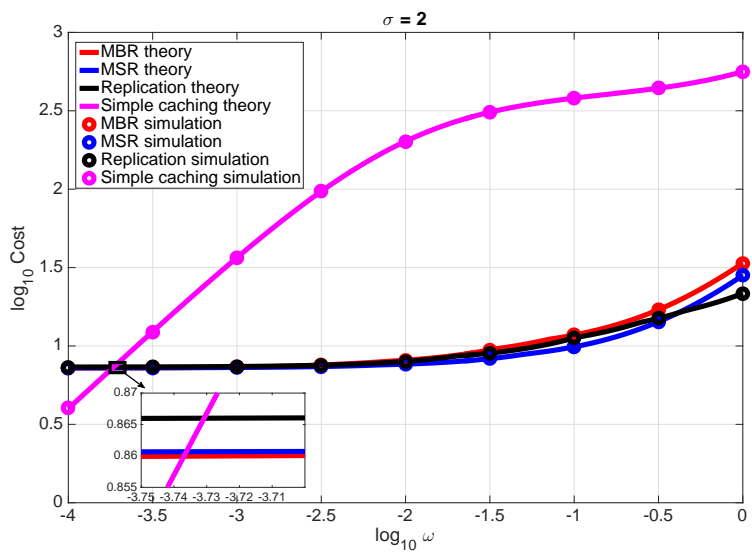}
	\caption{Costs versus file popularity for storage cost $\sigma=2$.}
	\label{fig:sim}
\end{figure}

In the setting of Figure~\ref{fig:sim}, each of the four caching methods becomes useful depending on the value of $\omega$. When the file popularity is low, maintaining redundancy wastes more transmission energy than is saved by D2D requests, and thus simple caching is preferred.

For a higher popularity, in the magnified range in the figure, we see that MBR coding is the optimal method as it has a lower repair bandwidth than MSR coding. This is where maintaining redundancy starts to pay off, and the cost function is dominated by the cost of repair as requests and thus file reconstructions are relatively rare compared to node failures.

When the popularity grows even larger, file requests become more abundant, and the cost of reconstruction starts dominating the cost function. Due to its low reconstruction bandwidth, MSR coding outperforms the other methods in this range of $\omega$. Finally, when the popularity is very high, replication yields the lowest cost. The reason why replication outperforms MSR, even though the reconstruction bandwidths are equal for both methods, is because replication only requires contacting the nearest storage node, while MSR requires contacting several nodes and the transmission energy cost increases proportionally to the fourth power of the distance.

Table \ref{tab:costsavings} shows example values on how much redundancy can decrease the cost compared to simple caching. We see that the performance gains are very notable for high file popularities.

\begin{table}[!h]
\begin{center}
    \begin{tabular}{| c | c | c |}
    \hline
    $\log_{10}\omega$ & Savings (\%) & Caching method \\ \hline
    -3.5 & 41.1 & \textcolor{red}{MBR} \\ \hline
    -3 & 80.1 & \textcolor{blue}{MSR} \\ \hline
    -2.5 & 92.3 & \textcolor{blue}{MSR} \\ \hline
    -2 & 96.2 & \textcolor{blue}{MSR} \\ \hline
    -1.5 & 97.3 & \textcolor{blue}{MSR} \\ \hline
    -1 & 97.4 & \textcolor{blue}{MSR} \\ \hline
    -0.5 & 96.7 & \textcolor{blue}{MSR} \\ \hline
    0 & 96.1 & Replication \\ \hline
    \end{tabular}
\end{center}
\caption{Cost savings compared to simple caching with the corresponding optimal coding methods and file popularities for $\sigma=2$, \emph{i.e.}, the setting in Figure~\ref{fig:sim}.}
\label{tab:costsavings}
\end{table}

\begin{figure}[!h]
\begin{subfigure}{0.5\textwidth}
\centering
\includegraphics[width=\textwidth]{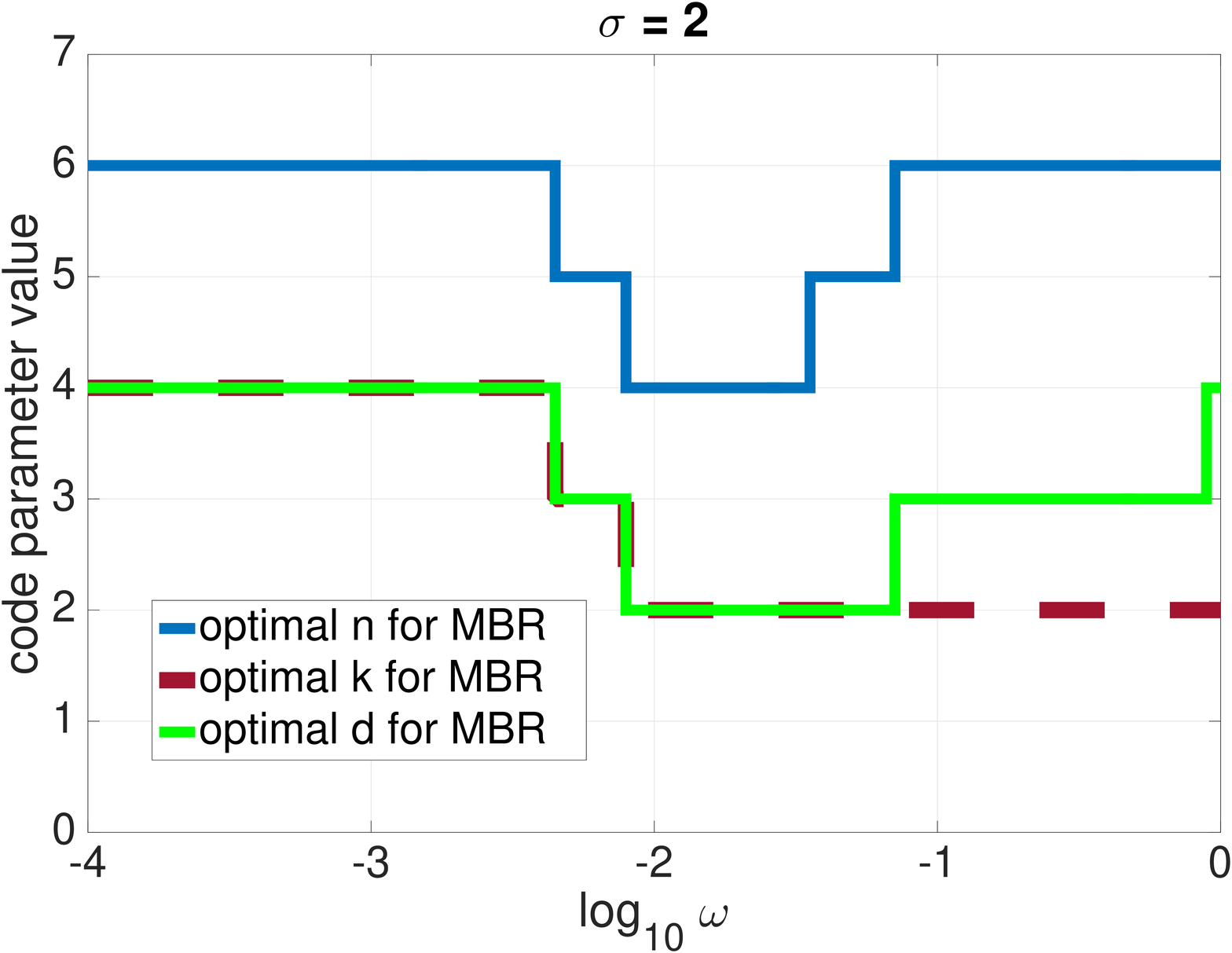}
\caption{Optimal MBR code parameters $(n,k,d)$.}
\label{fig:optMBR}
\end{subfigure}%
\begin{subfigure}{0.5\textwidth}
\centering
\includegraphics[width=\textwidth]{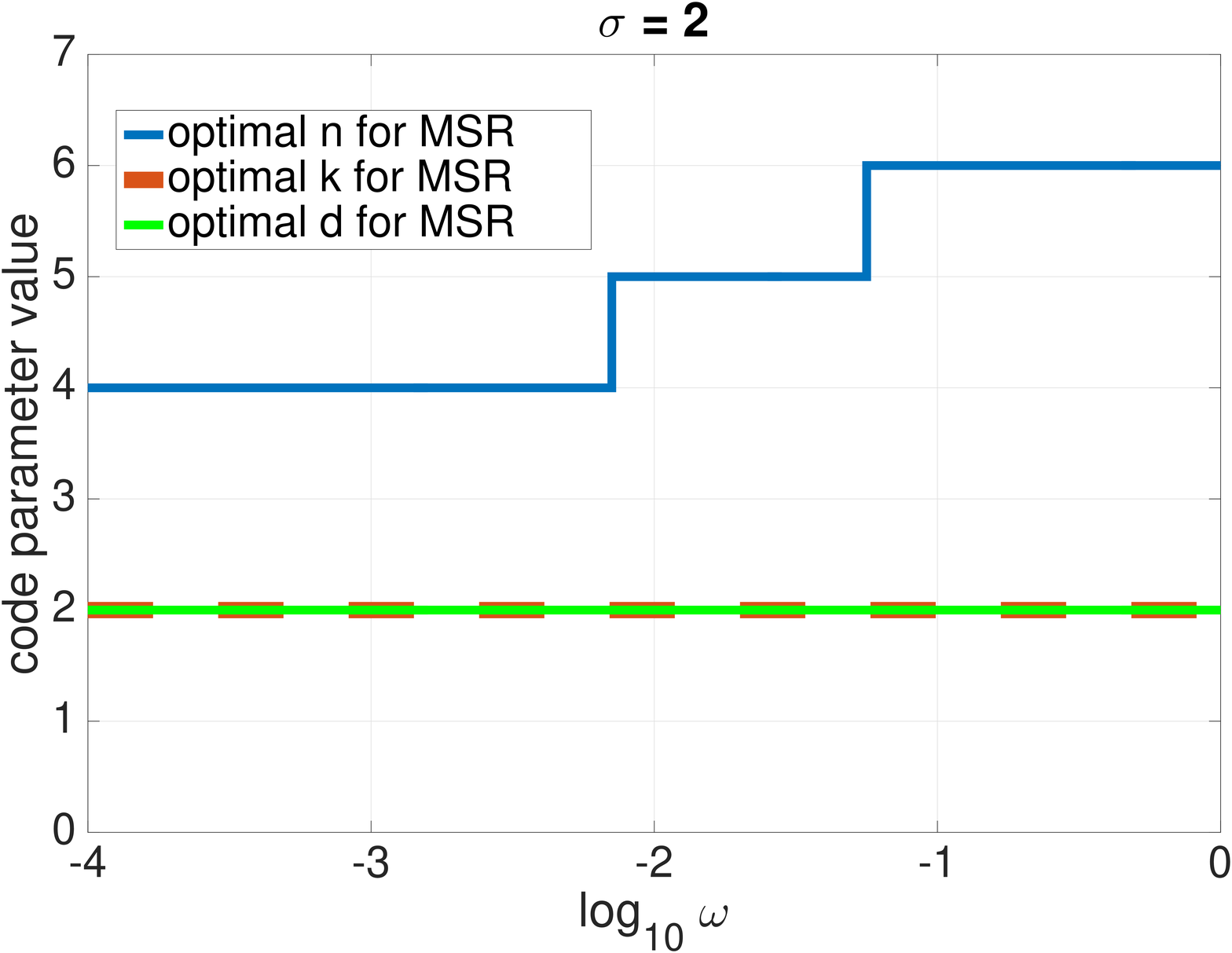}
\caption{Optimal MSR code parameters $(n,k,d)$.}
\label{fig:optMSR}
\end{subfigure}\\[1ex]
\begin{subfigure}{\textwidth}
\centering
\includegraphics[width=0.5\textwidth]{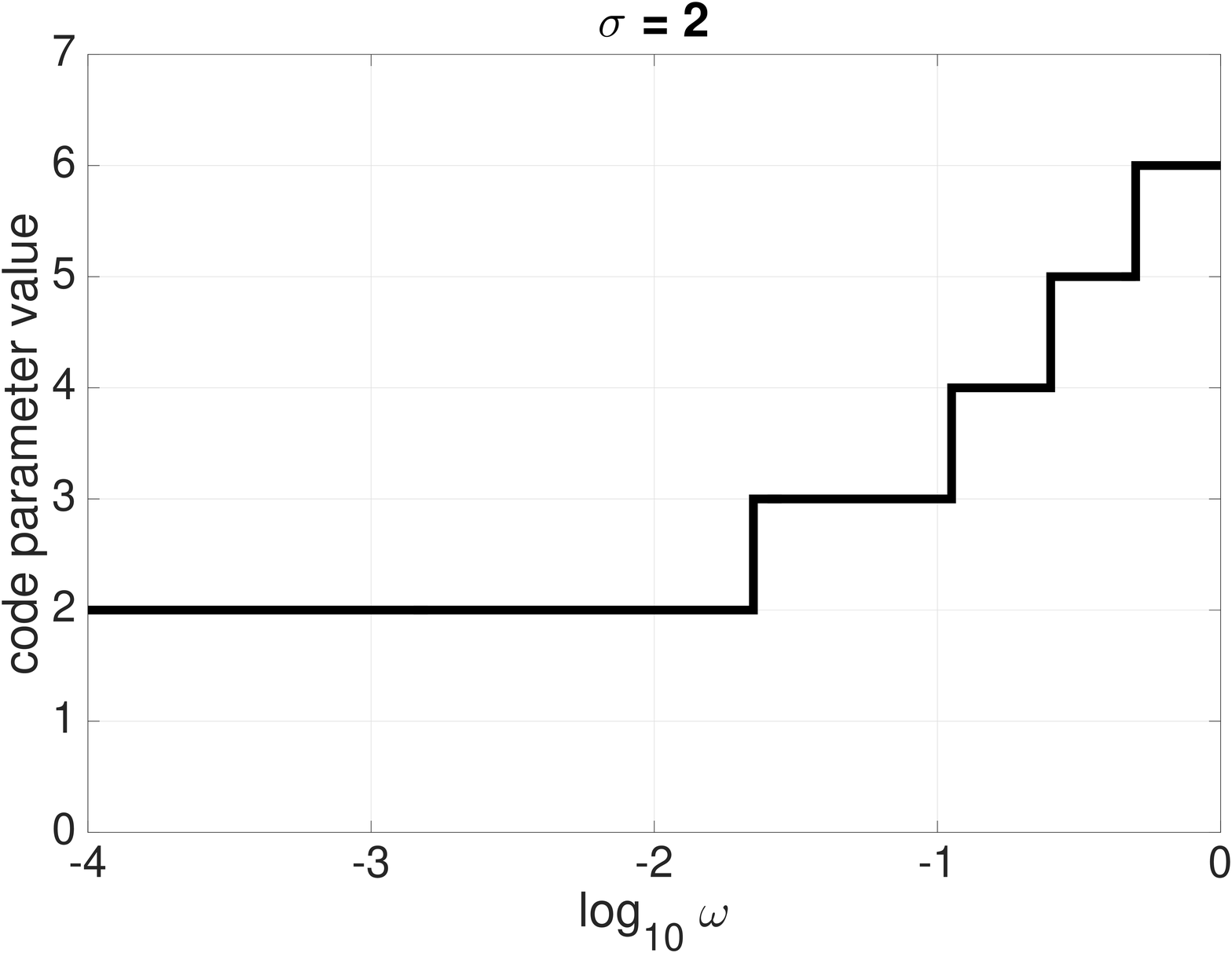}
\caption{Optimal number of replicas for replication $n$.}
\label{fig:optREP}
\end{subfigure}
\caption{Optimal parameters in the setting of Figure~\ref{fig:sim}.}
\label{fig:opt1}
\end{figure}

As previously mentioned, the code parameters $(n,k,d)$ for regenerating codes and $n$ for replication for the setting in Figure~\ref{fig:sim} were found through exhaustive searches, and they are depicted in Figure~\ref{fig:opt1}. In all cases, $n \in [2,6]$ for replication, and $n \in [3,6]$ for regenerating codes. It is a natural choice to have a relatively low upper bound for $n$ as $k$ and $d$ are upper-bounded by $n-1$, and it is impractical to establish a large number of simultaneous D2D connections.

In Figure~\ref{fig:optMBR}, we see the optimal code parameters for MBR coding. When the file popularity is low, both $k$ and $d$ are relatively high as the repair bandwidth ought to be low. We see an interesting dip of all parameter values approximately in the range $\omega \in (0.01,0.1)$. This is where the storage degree $n$ should be lowered to find an optimal balance between the number of failures and the reconstruction and repair degrees. Recall that the higher the storage degree, the higher the expected number of failures in a given time interval, while the lower the storage degree, the higher the expected transmission distances. Thus, finding the optimal $n$ is not trivial. If the popularity is high, it is the reconstruction energy that should be minimized. We see that $k$ should be low and $n$ should be high, which means that, for reconstruction, only the nearest storage nodes need to be contacted, and having a high density of storage nodes in turn implies short transmission distances.

In Figure~\ref{fig:optMSR}, the optimal code parameters for MSR coding are depicted. The repair degree $d$ and reconstruction degree $k$ should be kept low as this ensures that only the nearest storage node needs to be contacted. Also, it is more important for the failure rate to be low for relatively low popularities. When the popularity $\omega$ increases, so does the optimal storage degree $n$. This is again because the higher the storage node density in the cluster, the lower the expected transmission distances. Note that in this case the storage cost is relatively low and that a higher storage cost affects the behavior of the optimal parameter curves, as will be shown later in Figure~\ref{fig:MSRcodeparams}.

Lastly, Figure~\ref{fig:optREP} shows how the optimal storage degree $n$ for replication grows with increasing file popularity. As noted before, when the file popularity is high, it is important that the nearest replica is as close to the requesting user as possible -- thus the total number of replicas should be high, despite high total failure rates and storage costs.

\begin{figure}[!h]
\centering
	\includegraphics[width=.7\textwidth]{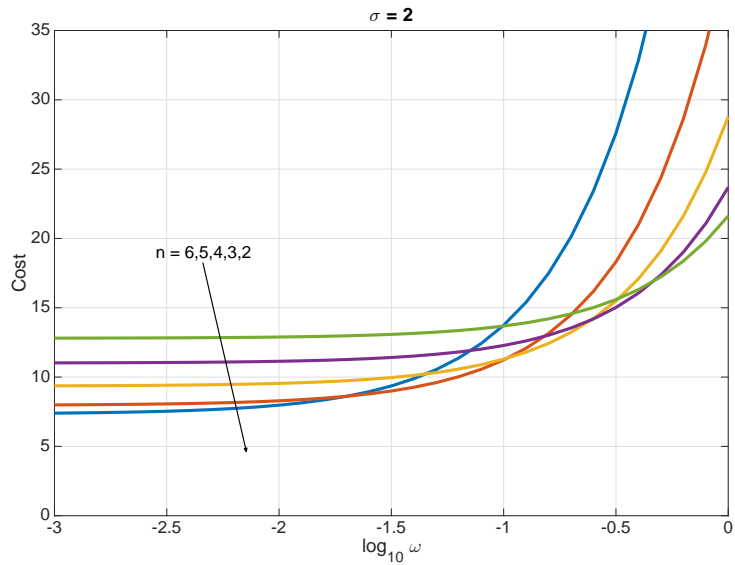}
	\caption{Cost of replication \eqref{repequ} versus file popularity for various storage degrees $n$. The lowest storage degree offers best results for low file popularities as the failure rate is low, but performs poorly for high popularities due to long transmission distances.}
	\label{fig:replication}
\end{figure}

We further illustrate the impact of the choice of the storage degree $n$ on the cost of replication in Figure~\ref{fig:replication}. We see that when the file popularity is low, it is unfavorable to have a high storage degree as this implies a high number of failures and thus plenty of upkeep and storage costs. To the contrary, when the popularity is high, it is important that the expected distance to the nearest caching node is short, and thus the number of replicas should be high. Based on these reasons, another interesting observation we make is that, despite assuming instant repairs, replication with one redundant copy, \emph{i.e.}, 2-replication, is not always the optimal method. Note that this is, as expected, a different result than that of \cite{globecom}, where we ignored distances and found that instant repair with 2-replication is optimal since it yields the lowest total failure rate.

In the previous setting, the storage cost $\sigma$ was chosen to be relatively low. Next, we investigate a case where the storage cost is much higher. This represents a case where there are plenty of potential files to be cached and it is important not to carelessly waste storage space so that as many files can be offloaded to the D2D community as possible.

\begin{figure}[!h]
\centering
	\includegraphics[width=.7\textwidth]{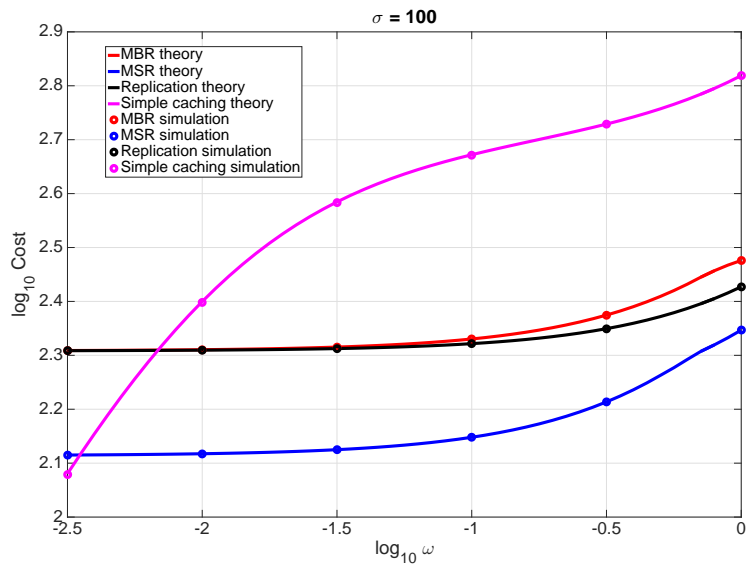}
	\caption{Costs versus file popularity for $\sigma=100$. When the file popularity and storage cost are high, MSR outperforms the other methods.}
	\label{fig:simMSRrocks}
\end{figure}

\begin{figure}
	\centering
	\includegraphics[width=.7\textwidth]{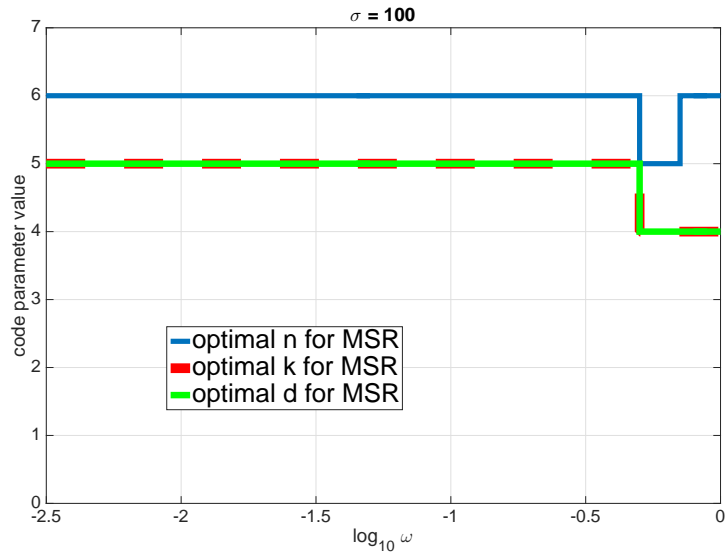}
	\caption{Optimal MSR code parameters $[n,k,d]$ for the case of Figure~\ref{fig:simMSRrocks}.}
	\label{fig:MSRcodeparams}
\end{figure}

In Figure~\ref{fig:simMSRrocks} we present cost versus file popularity for $\sigma=100$. We see that, as expected, MSR coding performs very well when the storage cost is high, as it is designed to minimize both the reconstruction bandwidth and the storage space consumption, while still maintaining a low repair bandwidth.
We underline that, although replication is a very simple method to add redundancy and it yields a low reconstruction cost as only the nearest caching node must be contacted, its drawback is wasteful storage space consumption. This leads to high storage costs, and consequently, MSR can yield a much lower total cost. This is illustrated in Table \ref{tab:costsavingsMSR} where we see significant cost savings when MSR is used as opposed to replication.
\begin{table}
	\begin{center}
    \begin{tabular}{| c | c |}
    \hline
    $\log_{10}\omega$ & Savings (\%) \\ \hline
    -2 & 35.8 \\ \hline
    -1.5 & 35.1  \\ \hline
    -1 & 33.0  \\ \hline
    -0.5 & 26.9 \\ \hline
    0 & 16.9 \\ \hline
    \end{tabular}
	\end{center}
	\caption{Cost savings by using MSR coding compared to replication with $\sigma=100$. When storage space is expensive, regenerating codes offer significant savings compared to naive replication.}
	\label{tab:costsavingsMSR}
\end{table}

The optimal values for the parameters of MSR with varying $\omega$ in the setting of Figure~\ref{fig:simMSRrocks} are shown in Figure~\ref{fig:MSRcodeparams}. Interestingly, the optimal choice is to set $d=k$. This implies that here \emph{traditional MDS coding is optimal}, which has been noted also in \cite{pedersen} for certain scenarios. Note that when $n=k+1$ and $d=k$ are the optimal parameters, which holds for most of the popularity values in Figure~\ref{fig:MSRcodeparams}, we can simply use the \emph{parity check code} with $k$ storage nodes and a single parity node. For very high file popularities, we see a similar dip in the optimal code parameters as in Figure \ref{fig:optMBR}, where both the reconstruction degree $k$ and the repair degree $d$ should be slightly lowered to obtain lower transmission distances for reconstruction and repair. 

\begin{figure}[!h]
\centering
	\includegraphics[width=.7\textwidth]{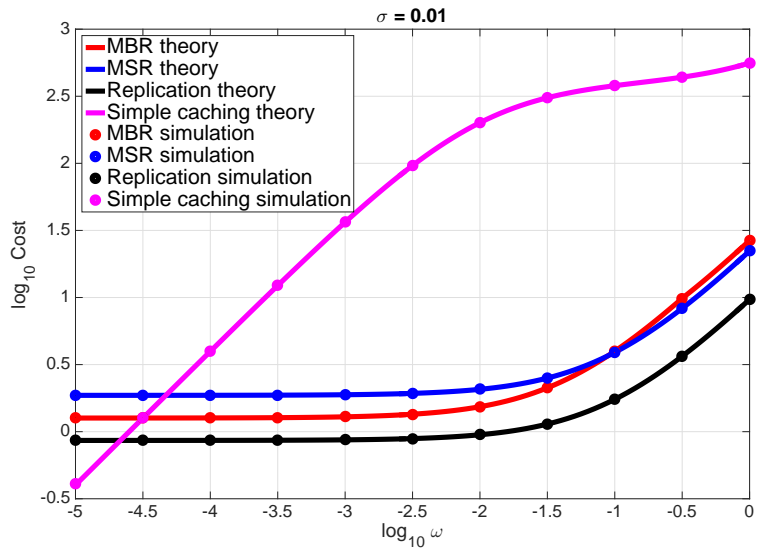}
	\caption{Costs versus file popularity for $\sigma=0.01$. When the storage cost is low, replication is preferred.}
	\label{fig:reprules}
\end{figure}

Finally, we assume a very low storage cost of $\sigma=0.01$ in Figure~\ref{fig:reprules}. As expected, replication is the preferred method. Numerical computations show that the optimal number of replicas is the maximum (six) for the whole popularity range considered. As the storage cost is very low, it is important to minimize the reconstruction and repair cost by minimizing the expected distance to the nearest caching node, that is, to fill up the cluster with as many replicas as possible.

\begin{figure}[!h]
\centering
	\includegraphics[width=.7\textwidth]{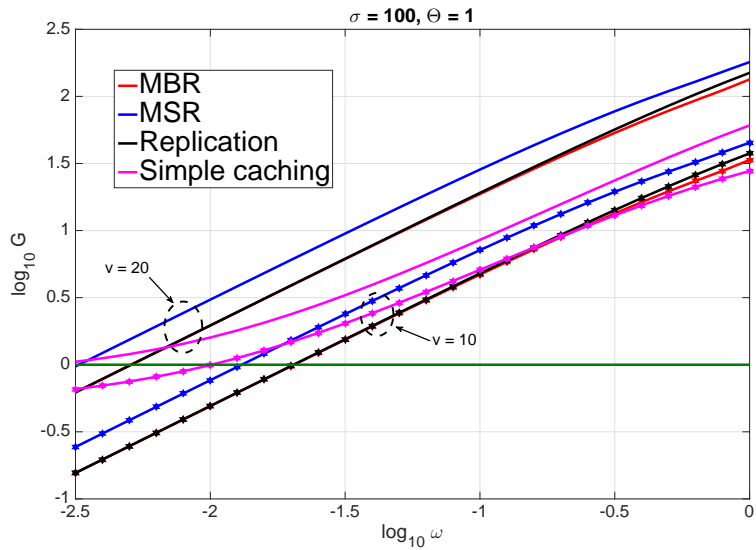}
	\caption{Theoretical operator gains $G$ \eqref{opegain} versus file popularity $\omega$ when $v=20$ (solid line) and $v=10$ (solid line with marker), with $\sigma=100$ and $\Theta=1$.}
	\label{fig:operight}
\end{figure}

We now turn our attention to the gains from an operator's point of view by plotting an example of \eqref{opegain}. In Figure~\ref{fig:operight}, we fix $\sigma = 100$ and $\Theta = 1$, and vary the base station distance ($v=10$ or $v=20$). These results present the ratio $G$ of the cost of only using the base station and the cost of caching the file using a given caching method. For example, when $\omega = 0.1$ and $v = 20$, using the base station is approximately $10^{1.5}\approx 32$ times more costly than maintaining a system using MSR coding.

The green line at zero in the figure is used as the reference to represents the case where downloading the file from the base station performs equally compared to a given caching method. Whenever a curve is above this reference line, it is beneficial to cache the file with the corresponding method. Here the storage cost is as high as in the setting of Figure~\ref{fig:simMSRrocks}.

We see that, for high file popularity, MSR outperforms the other methods. Although not depicted, the optimal MSR code parameters are $n=k+1=6$ and $d=k=5$, which again implies that the \emph{parity check code is optimal}. We also see that when the distance from the cluster to the base station increases from $10$ to $20$, there is approximately a lift of $0.5$ in the curves, which means that, roughly speaking, the operator gain increases threefold. The general trend is that the operator gain increases quickly with file popularity and base station distance, which implies that especially remote caching clusters can greatly benefit from D2D caching.

\section{Conclusions}
\label{sec:conclu}
We have studied the prospective benefits of distributed storage coding in a D2D caching
cluster, where communication cost grows with distance due to increasing pathloss. Our main objective has been optimizing the overall energy-efficiency of the network in terms of energy costs of communication and storage. We have found
that distributed storage coding can save more than 90\% in energy consumption compared to
caching without redundancy in a realistic scenario. Especially an error correcting code with
minimum storage overhead, here a parity check code or an MDS code, offers 
plenty of offload potential when storage space is moderately to highly expensive. However, when storage costs are low as compared to communication costs, simple repetition coding is preferred. When communication costs dominate over storage costs in the D2D community, physical proximity is of utmost
importance due to pathloss. Further, we have shown
that storage coding offers plenty of cost saving potential also from an operator's
point of view, part of which can be used to incentivize users to participate in a D2D caching community.

\section*{Acknowledgements}
The authors would like to thank Majid Gerami, Ejder Ba\c{s}tu\u{g}, Toni Ernvall, Pasi Lassila, and Lasse Leskel\"a for fruitful discussions.

\section{Appendix}
A summary of the parameters involved in the system model is given in Table~\ref{tab:parameters}. 

\renewcommand\arraystretch{1.25}
\begin{table}
\begin{tabular}{|c|l|l|}
  \hline
  \textbf{Parameter} & \multicolumn{2}{p{0.8\textwidth}|}{\textbf{Explanation}} \\
  \hline \hline
  $m$ &  \multicolumn{2}{p{0.8\textwidth}|}{\raggedright Expected total number of nodes in the cluster.} \\
  \hline
  $\lambda$ &  \multicolumn{2}{p{0.8\textwidth}|}{\raggedright Cluster passing rate of a single node, or node failure rate.} \\
  \hline
  $T$ &  \multicolumn{2}{p{0.8\textwidth}|}{\raggedright Expected time that a node stays active in the cluster, $T = 1/\lambda$.} \\
  \hline
  $\omega$ &  \multicolumn{2}{p{0.8\textwidth}|}{\raggedright File request rate of a single node. The expected file request interval is $1/\omega$.} \\
  \hline\hline
  $n$ &  \multicolumn{2}{p{0.8\textwidth}|}{\raggedright Storage degree: Number of nodes in the cluster storing a fragment or a replica of the file.}  \\
  \hline
  $k$ &  \multicolumn{2}{p{0.8\textwidth}|}{\raggedright Reconstruction degree: Number of nodes that must be contacted in order to reconstruct the entire file.} \\
  \hline
  $d$ &  \multicolumn{2}{p{0.8\textwidth}|}{\raggedright Repair degree: Number of nodes that a new, empty node must contact to repair the contents of a no longer available node.} \\
  \hline
  $\alpha$ &  \multicolumn{2}{p{0.8\textwidth}|}{\raggedright Number of data stored on each storage node. The total amount of data stored in the cluster is $n\alpha$, while the reconstruction bandwidth is $k\alpha$. For replication, $\alpha = 1$.} \\
  \hline
  $\beta$ &  \multicolumn{2}{p{0.8\textwidth}|}{\raggedright Amount of data traffic transmitted from a single storage node when repairing a lost node. The total repair bandwidth is $\gamma=d\beta$. For replication, $\beta = 1$.} \\
  \hline
  $\gamma$ &  \multicolumn{2}{p{0.8\textwidth}|}{\raggedright Total repair bandwidth $\gamma=d\beta$. For replication, $\gamma = 1$.} \\
 \hline  \hline
  $\sigma$ &  \multicolumn{2}{p{0.8\textwidth}|}{\raggedright Energy cost of storing a unit of data. In Section~\ref{subsec:operator} this is used to denote how much storing a single unit of data costs for the operator.} \\
  \hline
  $\Theta$ &  \multicolumn{2}{p{0.8\textwidth}|}{\raggedright Energy cost of wirelessly transmitting a unit of data over unit distance. This variable is used to weigh the cost of transmission energy consumed by the users as opposed to the weight $\sigma$ associated with the storage cost when calculating the operator gain in Section~\ref{subsec:savings_general}.} \\
    \hline
  $\Gamma$ &  \multicolumn{2}{p{0.8\textwidth}|}{\raggedright Pathloss exponent. Energy cost is modeled exactly as the $\Gamma^\text{th}$ power of the distance. The pathloss exponent of direct transmission between two nodes in the D2D cluster is $\GD$. The pathloss exponent of the downlink from the base station to a node in the cluster is $\GB$.} \\
  \hline
  $L_{r,\Gamma}(i,n)$ &  \multicolumn{2}{p{0.8\textwidth}|}{\raggedright Expected $\Gamma^\text{th}$ power of the distance from an arbitrary point in a disk of radius $r$ to its $i^\text{th}$ nearest neighbor out of $n$ nodes uniformly distributed in the disk. This is also the expected energy cost of transmitting a unit of data over the related distance.} \\
  \hline
  $\mathcal{E}_{\GB}(r,v)$ &  \multicolumn{2}{p{0.8\textwidth}|}{\raggedright Expected $\GB^{th}$ power of the distance between a node in the disk of radius $r$ and the base station located at a distance $v$ away from the center of the disk.} \\
  \hline
\end{tabular}
\caption{Notation summary.}
\label{tab:parameters}
\end{table}


\begin{thebibliography}{10}
\providecommand{\url}[1]{#1}
\csname url@samestyle\endcsname
\providecommand{\newblock}{\relax}
\providecommand{\bibinfo}[2]{#2}
\providecommand{\BIBentrySTDinterwordspacing}{\spaceskip=0pt\relax}
\providecommand{\BIBentryALTinterwordstretchfactor}{4}
\providecommand{\BIBentryALTinterwordspacing}{\spaceskip=\fontdimen2\font plus
\BIBentryALTinterwordstretchfactor\fontdimen3\font minus
  \fontdimen4\font\relax}
\providecommand{\BIBforeignlanguage}[2]{{%
\expandafter\ifx\csname l@#1\endcsname\relax
\typeout{** WARNING: IEEEtran.bst: No hyphenation pattern has been}%
\typeout{** loaded for the language `#1'. Using the pattern for}%
\typeout{** the default language instead.}%
\else
\language=\csname l@#1\endcsname
\fi
#2}}
\providecommand{\BIBdecl}{\relax}
\BIBdecl

\bibitem{globecom}
J. P\"a\"akk\"onen, C. Hollanti, and O. Tirkkonen, ``Device-to-Device Data Storage for Mobile Cellular Systems", in \emph{Proc. IEEE Global Communications Conference (GLOBECOM)}, pp.~671--676, Atlanta, GA, USA, December 2013.

\bibitem{macom}
J. P\"a\"akk\"onen, C. Hollanti, and O. Tirkkonen, ``Device-to-Device Data Storage with Regenerating Codes," in \emph{Proc. 8th International Workshop on Multiple Access Communications (MACOM)}, pp.~57--69, Espoo, Finland, September 2015.

\bibitem{cisco}
Cisco, ``Cisco Visual Networking Index: Global Mobile Data Traffic Forecast Update, 2015--2020," White Paper, http://goo.gl/l77HAJ, 2014.

\bibitem{janis}
P. J\"anis, C.-H. Yu, K. Doppler, C. Ribeiro, C. Wijting, K. Hugl, O. Tirkkonen, and V. Koivunen, ``Device-to-Device Communication Underlaying Cellular Communications Systems," in \emph{International Journal of Communication}, vol.~2, no.~3, pp.~169--178, June 2009.

\bibitem{doppler}
K. Doppler, M. P. Rinne, P. Janis, C. Ribeiro, and K. Hugl, ``Device-to-device Communications; Functional Prospects for LTE-advanced Networks," in \emph{IEEE International Conference on Communications Workshops, 2009}, pp.~1--6, Dresden, Germany, June 2009.

\bibitem{fodor}
N. Reider and G. Fodor, ``A Distributed Power Control and Mode Selection Algorithm for D2D Communications," in \emph{EURASIP Journal on Wireless Communications and Networking}, pp.~1--25, August 2012.

\bibitem{yu}
C.-H. Yu, O. Tirkkonen, K. Doppler, and C. Ribeiro, ``On the Performance of Device-to-Device Underlay Communication with Simple Power Control," in Proc. \emph{IEEE Information Theory Workshop (ITW)}, pp.~1--5, Seville, Spain, September 2013.

\bibitem{asadi}
A. Asadi, Q. Wang, and V. Mancuso, ``A Survey on Device-to-Device Communication in Cellular Networks," \emph{IEEE Communications Surveys \& Tutorials}, vol.~16, no.~4, pp.~1801--1819, Fourthquarter 2014.

\bibitem{maddah}
M. A. Maddah-Ali and U. Niesen, ``Fundamental Limits of Caching," \emph{IEEE Transactions on Information Theory}, vol.~60, no.~5, pp.~2856--2867, May 2014.

\bibitem{jeon}
S-W. Jeon, S-N. Hong, M. Ji, and G. Caire, ``Caching in Wireless Multihop Device-to-Device Networks," in \emph{Proc. IEEE International Conference on Communications (ICC)}, pp.~6732--6737, London, UK, April 2015.

\bibitem{gerami}
M. Gerami, X. Ming, and M. Skoglund, ``Partial Repair for Wireless Caching Networks With Broadcast Channels," \emph{IEEE Wireless Communications Letters}, vol.~4, no.~2, pp.~145--148, April 2015.

\bibitem{guocooperative} 
Y. Guo, L. Duan, and R. Zhang, ``Cooperative Local Caching and File Sharing under Heterogeneous File Preferences," arXiv:1510.04516, October 2015.

\bibitem{altman}
E. Altman, K. Avrachenkov, and J. Goseling, ``Distributed Storage in the Plane," in \emph{Proc. International Federation for Information Processing (IFIP) Networking Conference}, pp.~1--9, Trondheim, Norway, June 2014.

\bibitem{golrezaeibase}
N. Golrezaei, P. Mansourifard, A. F. Molisch, and A. G. Dimakis, ``Base Station Assisted Device-to-Device Communications for High-Throughput Wireless Video Networks," \emph{IEEE Transactions on Wireless Communications}, vol.~13, no.~7, pp.~3665--3676, July 2014.

\bibitem{bastug}
E. Ba\c{s}tu\u{g}, M. Bennis, and M. Debbah, ``Living on the Edge: The Role of Proactive Caching in 5G Wireless Networks," \emph{IEEE Communications Magazine}, vol.~52, no.~8, pp.~82--89, August 2014.

\bibitem{jidisse}
M. Ji, ``Fundamental Limits of Caching Networks: Turning Memory into Bandwidth," Doctoral dissertation, Faculty of the USC Graduate School, University of Southern California, 2015.

\bibitem{ott}
J. Ott and M. Pitk\"anen, ``DTN-based Content Storage and Retrieval," in \emph{Proc. IEEE WoWMoM Workshop on Autonomic and Opportunistic Communications (AOC)}, pp.~1--7, Espoo, Finland, June 2007.

\bibitem{lenders}
V. Lenders, G. Karlsson, and M. May, ``Wireless Ad Hoc Podcasting," in \emph{Proc. IEEE Conference on Sensor, Mesh, and Ad Hoc Communications and Networks (SECON)}, pp.~273--283, San Diego, CA, USA, June 2007.

\bibitem{jipaper}
M. Ji, G. Caire and A. Molisch. ``Fundamental Limits of Distributed Caching in D2D Wireless Networks," in \emph{Proc. IEEE Information Theory Workshop (ITW)}, pp.~1--5, Seville, Spain, September 2013.

\bibitem{chenenergy}
B. Chen and C. Yang, ``Energy Costs for Traffic Offloading by Cache-enabled D2D Communications," arXiv:1603.04660, March 2016.

\bibitem{gregori}
M. Gregori, J. Gomez-Vilardebo, J. Matamoros, and D. G\"und\"uz, ``Wireless Content Caching for Small Cell and D2D Networks," \emph{IEEE Journal on Selected Areas in Communications}, vol.~34, no.~5, pp.~1222--1234, May 2016.

\bibitem{afshang}
M. Afshang, H. S. Dhillon, and P. H. J. Chong, ``Fundamentals of Cluster-Centric Content Placement in Cache-Enabled Device-to-Device Networks," in \emph{Proc. IEEE Global Communications Conference (GLOBECOM)}, pp.~1--6, San Diego, CA, USA, December 2015.

\bibitem{pedersen}
J. Pedersen, A. Graell i Amat, I. Andriyanova, and F. Br\"annstr\"om, ``Repair Scheduling in Wireless Distributed Storage with D2D Communication," \emph{Information Theory Workshop (ITW)}, pp.~69--73, Jeju, South Korea, October 2015.

\bibitem{wangwu}
L. Wang, H. Wu, and Z. Han, ``Wireless Distributed Storage in Socially Enabled D2D Communications," \emph{IEEE Access}, March 2016, DOI: 10.1109/ACCESS.2016.2546685.

\bibitem{tanganalysis}
S. Tang and B. L. Mark, ``Analysis of Opportunistic Spectrum Sharing with Markovian Arrivals and Phase-Type Service," \emph{IEEE Transactions on Wireless Communications}, vol.~8, no.~6, pp.~3142--3150, June 2009.

\bibitem{hungrandom}
H.-N. Hung, P.-C. Lee, and Y.-B. Lin, ``Random Number Generation for Excess Life of Mobile User Residence Time," \emph{IEEE Transactions on Vehicular Technology}, vol.~55, no.~3, pp.~1045--1050, May 2006.

\bibitem{thaj}
S. Thajchayapong, ``Mobility Patterns in Microcellular Wireless Networks," \emph{IEEE Transactions on Mobile Computing}, vol.~5, no.~1, pp.~52--63, January 2006.

\bibitem{harrison}
P. Harrison and N. M. Patel, ``Performance Modelling of Communication Networks and Computer Architectures," \emph{International Computer Science Series}, Addison-Wesley, 1992.

\bibitem{dimakis}
A. G. Dimakis, P. B. Godfray, Y. Wu, M. J. Wainwright, and K. Ramchandran, ``Network Coding for Distributed Storage Systems," \emph{IEEE Transactions on Information Theory}, vol.~56, no.~9, pp.~4539--4551, September 2010.

\bibitem{rashmi}
K. V. Rashmi, N. B. Shah, and P. V. Kumar, ``Optimal Exact-Regenerating Codes for Distributed Storage at the MSR and MBR Points via a Product-Matrix Construction," \emph{IEEE Transactions on Information Theory}, vol.~57, no.~8, pp.~5227--5239, August 2011.

\bibitem{guodevice}
J. Guo, S. Durrani, X. Zhou, and H. Yanikomeroglu, ``Device-to-Device Communication Underlaying a Finite Cellular Network Region," arXiv:1510.03162, March 2016.

\bibitem{circlecircle}
E. W. Weisstein, ``Circle-Circle Intersection." From MathWorld -- A Wolfram Web Resource. http://mathworld.wolfram.com/Circle-CircleIntersection.html

\bibitem{srinidist}
S. Srinivasa and M. Haenggi, ``Distance Distributions in Finite Uniformly Random Networks: Theory and Applications," \emph{IEEE Transactions on Vehicular Technology}, vol. 59, no. 2, pp.~940--949, February 2010.

\bibitem{kendall}
A. Stuart and K. Ord, \emph{Kendall's Advanced Theory of Statistics}, vol.~1, pp.~351, Arnold, London, 1998.

\bibitem{wugame}
W. Wu, J. Lui, and R. T. Ma, ``A Game Theoretic Analysis on Incentive Mechanisms for Wireless Ad Hoc VOD Systems," in \emph{Proc. IEEE 10th International Symposium on Modeling and Optimization in Mobile, Ad Hoc and Wireless Networks (WiOpt)}, pp.~177--184, Paderborn, Germany, May 2012.

\bibitem{alogame}
F. Alotaibi, S. Hosny, H. El Gamal, and A. Eryilmaz, ``A Game Theoretic Approach to Content Trading in Proactive Wireless Networks," in \emph{Proc. IEEE International Symposium on Information Theory (ISIT)}, pp.~2216--2220, Hong Kong, June, 2015.

\bibitem{chencaching}
Z. Chen, Y. Liu, B. Zhou, and M. Tao, ``Caching Incentive Design in Wireless D2D Networks: A Stackelberg Game Approach," accepted to \emph{IEEE International Conference on Communications (ICC)}, Kuala Lumpur, Malaysia, May 2016.

\end{thebibliography}
\end{document}